\documentstyle[12pt]{article}
\newcounter{subeq}
\newcommand{\eqnreset}{\setcounter{subeq}{0}}

\newcommand{\Z}{\mbox{$Z^0$}}

\newcommand{\be}{\begin{equation}}
\newcommand{\ee}{\end{equation}}
\newcommand{\bea}{\begin{eqnarray}}
\newcommand{\eea}{\end{eqnarray}\eqnreset}
\begin{document}
%
%
\renewcommand{\theequation}{\arabic{equation}\alph{subeq}}
\renewcommand{\thefigure}{\arabic{figure}}
\renewcommand{\thetable}{\arabic{table}}
\renewcommand{\theenumi}{\roman{enumi}.)}
\renewcommand{\labelenumi}{\roman{enumi}.)}

\newcommand{\pngoes}{\raisebox{-3.0ex}{\mbox{
$\stackrel{\textstyle \longrightarrow}
{\stackrel{\scriptstyle | \vec{p}_i | >\!> m}
{\scriptstyle n\, {\rm large} } } \ $}}}
\def\be{\begin{equation}}
\def\ee{\end{equation}}
\def\Z{$Z^{\circ}$ }
\newcommand{\si}{$\sin^2 \Theta_W$}
\def \e{$|K_{L, S} (0) > $}
\def \h{$ H_{\rm eff}$}

\title{Dynamical Electroweak Breaking:  Issues and Challenges\thanks
{Invited talk at the ``Salamfest", held in honor of Abdus Salam at the
International Centre for Theoretical Physics, Trieste, Italy, March 1993.
To appear in the Proceedings of this conference. }}
\author{R .D.~Peccei \\
 {\footnotesize {Department of Physics, University of
   California, Los Angeles, CA 90024 }}}
\date{}

\begin{abstract}
After briefly remarking on alternatives for breaking the electroweak
symmetry, I discuss the implication that recent precision experiments at
LEP have for the symmetry breaking sector.  The difficulties associated
with generating fermion masses when the electroweak symmetry is broken
dynamically are exposed and an alternative to the walking technicolor -
extended technicolor scenario is suggested, based on models of
substructure.  Failures and lessons from trying to incorporate families
and mixing in these latter schemes are also discussed.
\end{abstract}
\maketitle
\thispagestyle{empty}

\addtocounter{page}{-1}
\newpage

It is a great pleasure and a singular honor for me to be talking at this
``Salamfest", particularly on a topic that is so closely connected to
one of Salam's central contributions to physics:  the theory of the
electroweak interactions.  The model for the electroweak interactions put
forth more than 25 years ago by Glashow, Salam and Weinberg \cite{GSW},
as more and more data was found to be in agreement with its predictions, has
made a transition from model, to theory, to paradigm.  LEP has provided
the last chapter in this saga of success.  The precise comparison of
LEP data with the predictions of the Glashow, Salam and
Weinberg theory provides
overwhelming evidence that the electroweak interactions are indeed
described by an $SU(2)\times U(1) $ gauge theory spontaneously broken to
$U(1)_{em}$, with some custodial global symmetry guaranteeing that $\rho
= 1$ \cite{Rolandi}. Remarkably, however, even though the $SU(2) \times
U(1)$ theory is well tested, the exact nature of the symmetry breaking
mechanism is still essentially unknown.  My remarks here will try to
address this issue and some of its ramifications, particularly
concerning dynamical symmetry breakdown.

As in any symmetry breakdown, the breakdown of the $SU(2)\times U(1)$
electroweak gauge symmetry to $U(1)_{em}$ is governed by an order
parameter.  What is uncertain, at the moment, is precisely what this
order parameter is.  Two alternative theoretical speculations exist
concerning the nature of this order parameter.  Either it is thought to
be:
\begin{description}
\item{i)} the vacuum expectation value (vev) of an elementary scalar
field $, < \Phi >$,
\end{description}
or it is assumed that
\begin{description}
\item{ii)} it is related to a dynamical condensate of fermion -
antifermion pairs, $< \bar{T} T>$, formed in a, as yet to be discovered,
underlying theory.
\end{description}
Both the physics and philosophy behind these two alternatives are quite
distinct. In particular, the first option is quite compatible with weak
coupling, while the second option necessarily requires strong coupling.

The simplest way to break $SU(2)\times U(1) \to U(1)_{em}$ is to
introduce a complex scalar doublet $\Phi$ into the electroweak theory,
giving this field an asymmetric potential

\be
V (\Phi) = \lambda (\Phi^{\dagger} \Phi - \frac{v^2}{2} )^2~~,
\ee
which leads to the breakdown.  The order parameter $< \Phi >$ is then
directly related to the scale $v \simeq (\sqrt{2} G_F)^{-1/2}
\simeq 250~ GeV$ introduced in the potential and the breakdown occurs
irrespective of the strength $\lambda$ of the potential.  In
supersymmetric versions of the Glashow, Salam and Weinberg model, where the
presence of scalar fields is natural \cite{Natural}, the quartic scalar
coefficients in fact are simply related to the $SU(2)$ and $U(1)$
couplings, so that in these cases these couplings are clearly weak
\cite{SUSY}.  Even without invoking supersymmetry, perturbative
control of the theory argues for a weak effective coupling at the scale
of the symmetry breakdown, $\lambda (v)$, so that the Landau pole in
this coupling constant evolution is beyond the Planck mass, where surely
new physics will enter \cite{MPP}. \footnote {In turn, weak coupling,
necessitates having a light Higgs scalar in the spectrum, since $M^2_H = 2
\lambda~ (v)~ v^2.$}

Just as scalar vevs are naturally connected with weak coupling, having
the $SU(2)\times U(1) \to U(1)_{em}$ order parameters be related to a
fermion-antifermion condensate, $< \bar{T} T>$, necessarily involves
strong coupling. This is particularly clear if the interactions which
gives rise to the fermion - antifermion condensates are those of a non
Abelian gauge theory, as in technicolor \cite{SW}.  The running coupling
constant squared of such theories $\alpha_T (q^2)$, just like that of
QCD, is weak at short distances but becomes strong as $q^2 $ decreases.
One can define a characteristic scale for these theories as the scale
where $\alpha_T~(q^2)$ becomes unity
\be
\alpha_T (\Lambda_T^2) = 1~~.
\ee
Condensate formation occurs precisely because the gauge interactions
become strong.  Thus the size of the order parameter $< \bar{T} T>$ will
be related to the scale where $\alpha_T~ (q^2)$ becomes strong, and one
expects
\be
< \bar{T} T>~\sim~ \Lambda^3_T~~.
\ee

The physics of the symmetry breakdown of the electroweak interactions is
most clearly manifested in the amplitudes for longitudinal gauge boson
scattering.  Thus, it is in $W_L W_L$ scattering where one may eventually
see a distinction between the weak coupling and strong coupling
alternatives discussed above.  At very high energies $E >> M_W$, it is
possible to establish a direct connection - through the, so called,
equivalence theorem of Cornwall, Levin and Tiktopolous \cite{CLT} -
between the amplitudes of $W_L W_L$ scattering and those of their
corresponding Goldstone bosons fields, $w$. \footnote{These are the
fields which are absorbed in the Higgs mechanism \cite{Higgs} and serve
to give the $W's$ mass.} Namely
\be
A_{W_L W_L} = A_{ww} + 0 ( \frac{M_W}{E})
\ee
Clearly in the case where the order parameter is the elementary doublet
vev $ <\Phi >$, because the scalar self coupling $\lambda$ is weak, the
Goldstone boson scattering amplitudes will also be weak and one is led
to expect weak interactions for the longitudinal gauge bosons.  On the
other hand, if the symmetry breakdown is dynamical, as a result of
condensate formation, the Goldstone bosons interactions in the
underlying theory are strong and thus one expects that there will be
strong interactions among the $W_L's$.

Even though the equivalence theorem is a high energy theorem, so that it
cannot be strictly applied near threshold, there should not be much
difference at low energy between the strongly coupled and weakly
coupled symmetry breaking sectors.  As is well known \cite{Low}, the
threshold behaviour of Goldstone boson interactions depends only on the
metric of the coset space of the breakdown and not on any other details.
Thus, independently of whether the Goldstone bosons $w$ are strongly or
weakly coupled, they have the same threshold interactions.
Consequently,
one expects that also the $W_L's$ should have the same interactions at
low energy, independently of exactly how the $SU(2)\times U(1)$ symmetry
is broken.

The distinction between the two symmetry breaking mechanisms discussed,
however, should become apparent at CM energies in the $W_L W_L$
subsystem of order $\hat{E}_{cm}~\simeq~ \sqrt{16 \pi v^2}~\simeq~
1.7~TeV$ \cite{LQT}.  At these energies, in the strongly coupled case
one expects resonance formation, but nothing particularly spectacular
for the weakly coupled case.  To get to these energies in the $W_L W_L$
subsystem one needs the CM energies of the LHC and SSC, because of the
energy degradation which occurs as one goes from protons, to quarks and,
finally, to $W's$.  Many studies have been done to see if signals of
strong $W_L W_L$ scattering, such as resonance formation, could be
visible at these future machines \cite{signals}.  The difficulty, of
course, is to dig out these signals from the background arising from
ordinary two-$W$ production.  Although after cuts there are not many
events left, in all cases examined signals of strong $W_L W_L$
scattering are detectable at SSC energies, with these signals being
somewhat more marginal for the LHC.

Although the issue of what triggers the $SU(2)\times U(1)$ breaking will
be eventually answered when the LHC and SSC become operational, it is
interesting to ask whether one can tell anything already now about this
question from the high precision electroweak data that has been gathered
at LEP.  LEP energies are quite ``low" to be really directly
sensitive to the
symmetry breaking sector, but how the symmetry is broken can influence
radiative corrections.
In the elementary scalar case,  as was
pointed out first by Veltman \cite{Veltman}, the dependence of the
radiative corrections on the Higgs scalar mass, which typifies the
symmetry breakdown, is given schematically by
\be
{\rm rad.~corr}.~\sim ~ \frac{\alpha}{\pi}~\ell n~ M_H/M_Z +
(\frac{\alpha}{\pi})^2~(\frac{M_H}{M_Z})^2~~.
\ee
Thus, at leading order, one feels only {\bf logarithmic} effects.
Quadratic dependence on the Higgs mass is only felt at $0(\alpha^2)$ and
is negligible if the Higgs scalar is light, as expected in this
scenario.

The effects of a possible strong coupling symmetry breaking sector could
be felt at LEP through modifications to the gauge propagators - the,
so called, {\bf
oblique corrections}\cite{Oblique}.  Since the typical scale
where these effects should begin to be
significant is of order $\sqrt{16 \pi v^2} ~>>~M_Z$, it is sensible to
expand the vacuum polarization tensors for the gauge fields in a power
series in $q^2$, retaining just the first two terms \cite{PTABKL}:
\be
\Pi_{AB}~(q^2) = \Pi_{AB} (0) + q^2 \Pi^{\prime}_{AB}~(0)
\ee
Here the pair $\{AB\}$ spans the 4 possible gauge field configurations
$\{A B  = WW; Z Z; \gamma \gamma; \gamma Z\}$.
Electromagnetic gauge invariance implies that $\Pi_{\gamma \gamma}~(0) =
\Pi_{\gamma Z}~(0) = 0$, so that in toto one has 6 possible
constants characterizing the vacuum polarization tensors in this
approximation.  Since one is assuming that the electroweak theory is
$SU(2)\times U(1)$, three combinations of these constants can be fixed
in terms of the three independent parameters which typify this theory,
the two coupling constants $g_1$ and $g_2$ and $v$. \footnote{More
practically, instead of~\protect$g_1, g_2$ and \protect$v$
one uses three other related
parameters which are more accurately
known experimentally:~\protect$\alpha, M_Z$
and~\protect$G_F$, the Fermi constant measured
in~\protect$\mu$ decay.}  The other three
combination of parameters can in principle be determined from experiment
and then compared to what is expected by different schemes for symmetry
breakdown.

A standard set of parameters, denoted by $S, T$ and $U$ by Peskin and
Takeuchi \cite{PT} and by other equivalent symbols by other authors, has
emerged as being the most convenient to perform this analysis.  It
turns out
that, of the three, only S is moderately sensitive to the symmetry
breaking sector, because $T$ and $U$ are dominated by other
uncertainties. \footnote{\protect$T$ depends quadratically on the still
unknown top quark mass and so its measured value serves mostly to
constrain $m_t$.  $U$, on the other hand, depends sensitively on the
value of the $W$ mass and the present experimental uncertainty in this
value precludes using $U$ as a sensitive probe of the symmetry breaking
sector.} Furthermore, if the symmetry breaking sector conserves an
approximate vectorial $SU(2)$ symmetry - as it surely must, since
experimentally the $\rho$  parameter is very near unity - one can write
$S$ directly in terms of the vacuum polarization tensors for $SU(2)_V$
and $SU(2)_A$ currents, related to the $SU(2)\times U(1)$ currents,
facilitating in this way the comparison
with models of dynamical symmetry breaking.  One finds \cite{PTABKL}
\be S = - 4 \pi \{ \Pi^{\prime}_{VV} (0) - \Pi^{\prime}_{AA} (0) \} =
\int^{\infty}_0 \frac{ds}{s} [ v(s) - a(s)]~~,
\ee
where the second line makes use of a spectral decomposition for the
vector and axial vector current correlation functions.

One can extract what values of $S$ (and $T$ ) are allowed by comparing
electroweak observables [like the hadronic width of the $Z$] to the
theoretical predictions.  To do so one needs some reference value of
$m_t$ and~$ M_H$ to fix the expectations of the standard model, with
elementary scalar field breaking \footnote{The dependence of the results
on a standard model reference point can actually be avoided \cite{ABC}.
This dependence is, at any rate, very mild, as can be seen from Fig. 1.}.
That is, for each experimental observable one has formulas of the type
\cite{PT}
\be
O_{\rm exp}~=~ 0_{sm}~ + \alpha_0~ S + \beta_0~ T + \gamma_0 U
\ee
with $\alpha_0, \beta_0$ and $\gamma_0$ computable coefficients [e.g.
for $\Gamma_Z$, using as reference point $M_H = 1~TeV, m_t = 150~GeV$,
one has $(\Gamma_Z)_{\rm exp}~=~ 2.4 87~\pm~ 0.010~GeV~~;~~$\linebreak
$(\Gamma_Z)_{sm}~=~ 2.4 84~GeV; \alpha_0~=~ - 9.58 \times
10^{-3}~GeV~~;~~ \beta_0 = + 2.615 \times 10^{-2}~GeV~~;~~\gamma_0 = 0]$.
A global fit of all precision electroweak data, performed by
Peskin and Takeuchi \cite{PT} is shown in Fig. 1.  As can be seen, the
combined analysis of all data favors negative values of $S$ with
\be
S~\simeq~ - 0~(1)~~.
\ee

\begin{figure}[h]
\vspace {10cm}
\caption[]{ 68\% C.L. and 90\% C.L. curves in the \protect$S$ and
\protect $T$ plane determined by the global fit to all data carried out
by Peskin and Takeuchi \protect\cite{PT}. Very similar
results have also been obtained by \cite{KL} \cite{ABJ}.
In the figure the predictions of the standard model and of an $SU(4)$
technicolor theory with $n = 1$ and $n = 4$ are shown, as a function of
$m_t$.  The crosses in the figure denote $m_t$ values which, starting
from the bottom, increase by $20~GeV$ from $m_t = 90~GeV$.}
\end{figure}

This value can be compared to the expectations of various dynamical
symmetry breaking models.  The simplest of these models is technicolor
\cite{SW}.  Here one imagines that the condensates which break
$SU(2)\times U(1)_{em}$ are those of an underlying $SU(N)_T$ theory,
whose technifermions come in $n$ replications of left-handed doublets
$(\begin{array}{c}
U_i\\
D_i \end{array})_L, i= 1,2, \cdots, n$ and right-handed singlets $U_{iR},
D_{iR}$ - much as ordinary quarks do.
The condensates
\be
< \bar{U}_i U_j > ~= ~< {\bar{D}}_i D_j > ~=~ \delta_{ij} \Lambda^3_T
\ee
then break $SU(2) \times U(1) \to U(1)_{em}$ precisely as $n$ replicas
of scalar doublets $\Phi_i$ would.  Because this theory is just a
``scaled up" version of QCD, with the dynamical scale $\Lambda_T~\sim~ v
{}~\sim~ 250~$GeV, one can estimate the parameter $S$ from QCD.  By
dominating the vector and axial vector spectral functions in QCD by the
$\rho$ and $A_1$ poles one deduces that \cite{PT}
\be
S_{QCD} = \frac{6 \pi f_{\pi}^2}{m_{\rho}^2}
\ee
and hence for the $SU(N)_{T}$ theory one has
\be
S~ \simeq ~n \frac{N}{3} S_{QCD}~ \simeq~ n \frac{N}{3} (\frac{6\pi
f_{\pi}^2}{m_{\rho}^2})~\simeq~ 0.08~ nN
\ee
The vertical lines in Fig. 1 represent the expectations of such a simple
technicolor model with $N = 4$ and $n = 1$ or 4 where, however, the QCD
spectral integral was computed out without resorting to $\rho$ and $A_1$
dominance \cite{PT} \footnote{These results agree quite well, however,
with those obtained by assuming simple $\rho$ and $A_1$ dominance.}.  As
can be seen from this figure, the theoretically
``more realistic" $n = 4$ technicolor
model is about 3$\sigma$~ away from the experimentally allowed region.

Although this result is not very encouraging for dynamical symmetry
breaking theories, I do not believe it represents a death knell for these
theories.  First of all, one can argue, rather convincingly, that in
more sophisticated theories with dynamical symmetry breaking of $SU(2)
\times U(1)$ - the so called, walking technicolor theories \cite{WTC}, where
$\alpha_{T} (q^2)$ runs slowly - that the integral over the spectral
functions in $S$ should converge much more slowly, so that $S$ should be
below the estimate given above \cite{TT}.  Furthermore, if one
worries about how
fermions get mass in theories where $SU(2)\times U(1)$ is broken
dynamically, it becomes quite clear that a naive scaling up of QCD will
just not do!  I turn to discuss this important point further.

Simple technicolor models do not address the issue of fermion masses
unless one introduces some communication between ordinary fermions, $f$
- the quarks and leptons - and the fermions which condense, $T$.  This is
done ordinarily by introducing a second underlying theory - a, so
called,
extended technicolor theory (ETC) \cite{ETC} - to connect the $f$ and
$T$ fermions together.  If these fermions sit in the same ETC
representation and the ETC theory is a spontaneously broken gauge
theory, broken at a large scale $\Lambda_{ETC}$, one generates in this
way an effective four-fermion interaction

\be
{\cal{L}}^{ETC}_{eff}~=~ \frac{1}{\Lambda^2_{ETC}}~ ({\bar{T}}_L T_R)
({\bar{f}}_L f_R)~~.
\ee
This term, when the $T$ fermions condense $< {\bar{T}}_L T_R > \sim
\Lambda_T^3$~~,
gives the fermions $f$ a mass
\be
m_f~\sim~  \frac{<{\bar{T}}_R T_L>}{\Lambda^2_{ETC}}~\sim~
\frac{\Lambda^3_{T}}{\Lambda^2_{ETC}}
\ee

Given that the technicolor scale $\Lambda_{T}~\sim~ v$, the above
formula has a generic difficulty.  If one wants to generate with it the
top quark mass, because top is so heavy, the scale $\Lambda_{ETC}$
cannot be too large.  However, if $\Lambda_{ETC}$ is only in the $(1
-10)$ TeV range, then one cannot avoid large flavor changing neutral
current interactions (FCNC), which are not observed experimentally
\cite{DE}.  If $\Lambda_{ETC}$ is large, to avoid the FCNC problem, then
one can never generate in this way large enough fermion masses.

Although the above conundrum is not the only problem of ETC theories,
this problem can be ameliorated by walking technicolor models (WTC)
\cite{WTC}, which have more realistic dynamics.  Furthermore, WTC
theories can also ameliorate some of the other endemic problems of ETC,
like having too light pseudo Goldstone bosons.  So, if one believes that
the symmetry breakdown of $SU(2) \times U(1)$ to $U(1)_{em}$ is
dynamical, it is much more sensible to focus on WTC models.  Let me
briefly indicate what is the underlying idea in these models.  The
identification of $\Lambda_{T}$, the scale of the condensate, with $v$,
the scale related to the $W$ mass, is borrowed from QCD.  In QCD,
indeed, both the scale which measures the size of the quark condensate
$< \bar{u} u >$ - which triggers chiral symmetry breakdown -
and that which measures the pion decay constant $f_{\pi}$
- associated with the coupling of the broken chiral currents to the
Goldstone pion fields - are the same.  That is
\be
< \bar{u} u > \sim f_{\pi}^3~\sim~ \Lambda^3_{QCD}
\ee
In WTC models, it turns out that the techniquark condensates $< \bar{T}
T>$ have a scale which is much bigger than $v^3$. Hence, one can
obtain rather large masses for sizable ETC scales $\Lambda_{ETC}$,
avoiding the FCNC $\leftrightarrow m_t$ conundrum.

Both $< \bar{T} T>$ and $v$, the technipion decay constant, are related
to the techniquark self energy $\Sigma (p)$, but probe different momentum
scales in this function.  Graphically, the relation of $< \bar{T} T>$
and $v^2$ to $\Sigma (p)$ can be easily inferred from Fig. 2.  Retaining
only the leading momentum behaviour - neglecting logarithmic terms - one
has \cite{Holdom}
\[< \bar{T} T >~ \sim~ \int \frac{d^4 p}{p^2} \Sigma (p) \sim \int
d p^2 \Sigma (p)~~,\]
{}~~~while
\begin{equation}
v^2~ \sim~ \int~ \frac{d^4 p} {(p^2)^2} \Sigma^2 (p) \sim \int
\frac{d p^2} {p^2} \Sigma^2 (p)~~.
\end{equation}

For an asymptotically free theory the chirality breaking self energy
$\Sigma (p)$ falls at large momentum as \cite {LP}
\be
\Sigma (p) \sim \frac{\Sigma (0)^3}{p^2}~~,
\ee
where again I have neglected logarithmic factors.  Here $\Sigma (0)$
serves as the order parameter for the breakdown of the global chiral
symmetries of the theory.  Thus, one expects that it be of order of the
dynamical scale of the technicolor theory:  $\Sigma (0)~\sim~
\Lambda_T$.

Because the integral over $\Sigma (p)$ which enters for $v^2$ is
(essentially) convergent, one sees that $v^2$ probes the region of the
techniquark self energy near $p^2 = 0$,
\be v \sim \Sigma (0)\sim \Lambda_{T}~~.
\ee
On the other hand, the condensate $< \bar{T} T>$ feels much more the
large momentum structure of $\Sigma (p)$.  In particular, the region of
$p^2$ important for fermion mass generation is $p^2 \sim
\Lambda^2_{ETC}$ and one has
\be
m_f \sim \frac{< \bar{T} T>}{\Lambda^2_{ETC}} \sim
\frac{1}{\Lambda^2_{ETC}} \int^{\Lambda^2_{ETC}} d p^2 \Sigma (p^2)
\simeq \Sigma (\Lambda_{ETC})~~.
\ee
\newpage
\begin{figure}[h]
\vspace{6cm}
\caption[]{Graphs relating \protect$< \bar{T} T>~(a)$ and
\protect $v^2~(b)$ to the techniquark self energy \protect$\Sigma (p)$.}
\end{figure}

The value of $\Sigma (\Lambda_{ETC})$ is not known a priory, without a
dynamical calculation.  If one assumes that at this scale the
technifermion self energy already  has achieved its asymptotic
value $\Sigma (p) \sim
\Sigma (0)^3 /p^2$, then indeed one reproduces for $m_f$ the naive
estimate of before:
\be
m_f \sim \Sigma (\Lambda_{ETC} ) \sim \frac{\Sigma
(0)^3}{\Lambda^2_{ETC}} \sim \frac{\Lambda^3_{T}}{\Lambda_{ETC}^2}~~.
\ee
The assumption one makes in walking technicolor models is that the
dynamics is such that at $\Lambda_{ETC}, \Sigma (\Lambda_{ETC})$ is
still much above its ultimate asymptotic value.  Thus one can have large
masses for fermions even for large values of $\Lambda_{ETC}$, solving the
FCNC $\leftrightarrow m_t$ conundrum.

To obtain a behaviour of the type
\be
\Sigma (\Lambda_{ETC})~>>~ \frac{\Sigma(0)^3}{\Lambda^2_{ETC}}
\ee
it is necessary that all physical quantities evolve slowly with momentum
- hence the moniker walking for these theories.  This can be achieved if
the underlying theory is very  nearly not asymptotically free or,
perhaps, a fixed point theory \cite{WTC}.  Furthermore, if at
$\Lambda_{ETC}$ the coupling constant of the walking technicolor theory
is still rather strong, so that $\Sigma (p)$ does not take yet its
perturbative asymptotic value, it is clear that one cannot simply
decouple the ETC and the WTC dynamics from each other.  The presence of
the ETC interactions can therefore influence the self energy functions for
different technifermions differently, which may provide a physical
rationale for some of the observed hierarchies in the quark and lepton
mass matrices \cite{H}.

Rather than pursue further ETC/WTC theories here, I would like to
indicate an alternative possibility in which also the dynamics of fermion
mass generation and that of $SU(2) \times U(1)$ breaking are naturally
interlocked.  This possibility is realized in composite technicolor
models, where both the light fermions we see (the quarks and leptons) and
the fermions which condense (the technifermions) are bound states of some
fundamental preon theory.  In these kinds of theories, the preon dynamics
produces effective 4-fermion interactions involving technifermions and
light fermions which are analogous to the, perhaps more familiar, ETC
interactions.  The formation of technifermion condensates, combined with
the presence of these interactions, then allows mass to be generated for
the light fermions.

In what follows I want to describe a toy attempt in this direction,
developed in collaboration with S. Khlebnikov \cite{KP}.  Our model
produces one generation of quarks, say $t$ and $b$, as preon bound
states  \footnote{Leptons are not included in the model, but there are
no hypercharge anomalies since these are cancelled by the presence of
the techniquark bound states in the spectrum.}.  These states are almost
point-like, since their size is much smaller than their Compton's
wavelength
\be
< r_q >~<<~\frac{1}{m_q}~~.
\ee
However, one obtains a top-bottom mass hierarchy precisely because top
is more extended than bottom,
\be
\frac{m_t}{m_b}~\sim~ \frac{< r_t>^2}{<r_b>^2}~~.
\ee
The root cause for quark mass generation in the model is the
formation of certain $SU(2)\times U(1)$ breaking condensates
of other fermionic preon bound states, which act precisely as
techniquarks.  Thus, in the model, the $W$ and $Z$ get mass dynamically
by the same condensates which also generate quark masses.  As it will
become clear below, the model is rather uneconomical.  However, it is an
interesting toy laboratory in which to study a number of dynamical
issues.  Furthermore, the difficulties one encounters in trying to
incorporate families in extensions of this model
are also quite illustrative.

The model is based on a chiral gauged preon model.  One can argue
that certain of the global chiral symmetries in the model are preserved
in the binding.  As a result of these global chiral symmetries, there are
a number of massless bound states $B$ in the spectrum.  These states,
however, acquire mass when one gauges some subset of the preons in a
vector-like manner, as a result of condensate formation. After this
further gauging,
the original
massless bound states $B$ split into three different set of states
\be
B = \{ q, T, M \}~~,
\ee
with $q$ being the quarks $(t, b),~ T$ being techniquarks and $M$ being
megaquarks.  The metacolor gauge interaction at the preon level
leads to the formation of $< \bar{M} M>$ condensates which, in turn,
serve to produce effective ETC interactions between the quarks and
techniquarks. The technicolor gauge interaction then, through the
formation of $< \bar{T} T>$ condensates, leads to the appearance of the
$W$ and $Z$ masses and of mass terms for the quarks - these latter
masses originating as a result of the effective ETC interactions.

In more detail, the preon dynamics of the model we considered is based
on replicas of an $SU(6)$ preon theory with 10 Weyl preons in the
fundamental representation, $F_a (a = 1, \cdots, 10)$, and one preon in the
symmetric conjugate representation, $\bar{S}$.  Such a theory is chiral,
but has no gauge anomalies.  It has a nominal chiral global symmetry
\be
G = SU(10) \times U(1)_Q
\ee
where $U(1)_Q$ is the anomaly free combination of $F$ and $\bar{S}$
fermion number.  However, one can argue dynamically \cite{BY} that a
smaller symmetry than $G$ is preserved in the binding, namely
\be
H = SU(6) \times SU(4)\times U(1)_Q^{\prime}~~.
\ee
Furthermore, one can identify the set of bound states $B$ which are
massless by matching the $H$ anomalies at the preon level with those at
the bound state level \cite{Hooft}.  These states are readily seen to
comprise two different kinds of states
\begin{eqnarray}
B = \left\{\begin{array}{cc}
B_1  \sim & \mbox{(15; 1, -5/3)} \\
B_2  \sim & \mbox{(6; 4; -5/6)} \end{array} \right.
\end{eqnarray}
where, in the above, I have given the transformation of these
states under $H$.

The bound states $B$ feel effective interactions
among each other, as a result of their common underlying preonic
structure \footnote{ The states $B$ have the following schematic
preonic structure:~\protect$B_{ab} \sim F^T_a \sigma_2 \sigma_{\mu} F_b
\sigma^{\mu} \bar{S}$.}.  There are $H$ invariant dimension 6
interactions, scaling as $\Lambda^{-2}_c$,  with $\Lambda_c$ being the
dynamical scale of the preon theory,
\be
{\cal{L}}_9 = \frac{1}{\Lambda^2_c} (\bar{B} \gamma^{\mu} \lambda B) (
\bar{B} \gamma_{\mu} \lambda B)
\ee
and dimension 9 interactions, scaling as $\Lambda_c^{-5}$
\be
{\cal{L}}_9 = \frac{1}{\Lambda_c^5}~ B_1 B_1 {\bar{B}}_2 {\bar{B}}_2
{\bar{B}}_2 {\bar{B}}_2~~.
\ee
These latter interactions are not so important in the one generation model
under study, but they play an important role in multigeneration models,
since they violate individual $B_1$ and $B_2$ number.

The actual model studied in \cite{KP} is based on 3 replicas of these
$SU(6)$ preon models, with the individual models constructed so that
among their $B$ states one generates, respectively, a $t_R$ state, a
$b_R$ state and the $(\begin{array}{c}
t \\
b\end{array} )_L$ doublet of states.  The first two $SU(6)$ models
are identical in content with their 10 (right-handed) Weyl preons
feeling different gauge interactions.  One gauges an
$SU(3)_c \times SU(3)_T\times SU(4)_M$ group of color, technicolor and
metacolor interactions at the preon level, with the $10~ F_a$ preons
transforming under this group as
\be
F_a = \{ ( \bar{3}, 1, 1) \oplus (1, \bar{3}, 1) \oplus (1,1,4)\}~~.
\ee
The final $SU(6)$ preon theory has a doubled set of preons - $20~
F_a$ and $2~\bar{S}$ - so as to be able to introduce at the preon level
the electroweak $SU(2)\times U(1)$ interactions.  These preons are
(left-handed) Weyl states and are organized in
doublets of $SU(2)$, except for the
preons which feel the metacolor interactions, which are $SU(2)$ singlet
states.  Thus, in
the last preon theory, the preons transform as \footnote{All
preons also have appropriate \protect$U(1)$ quantum
number\protect\cite{KP}.}
\be
F_a = \{ (\bar{3}, 1, 1, 2) \oplus (1,\bar{3}, 1, 2) \oplus 2 (1, 1, 4, 1)
\};~ \bar{S} = (1, 1, 1, 2)~~.
\ee

The $B_1$ and $B_2$ massless bound states of each of these $3~ SU(6)$
preon theories can be classified immediately in terms of $SU(3)_c\times
SU(3)_T\times SU(4)_M$.  Since the antisymmetric combination of two
${\bar{3}}'s$ is a 3, one has
\[ B_1 \sim \{ 3,1, 1) \oplus (1,3,1) \oplus
(\bar{3}, \bar{3}, 1) \}\]
and
\be B_2 \sim \{ (\bar{3}, 1, 4) \oplus (1,
\bar{3}, 4) \}
\ee
Thus the $B_1$ states contains both quarks and techniquarks, while the
$B_2$ states are the only massless bound states which feel metacolor.

Gauging the metacolor group will cause the formation of condensates of
the $B_2$ states, produced by the right-handed $SU(6)$ preon theories,
with those produced by the left-handed $SU(6)$ preon theory.  These
condensates tie the left and right theories together, but preserve $SU(2)
\times U(1)$.  Specifically, one has two such condensates forming,
\footnote{The notation employed here is to append a superscript \protect
$t_R, b_R$ or $L$ to indicate to which preon theory the bound states
belong.  Note that there are 2 \protect~ $B^L_2$ bound states formed,
denoted by \protect$B^{1L}_2$ and~ \protect$B^{2L}_2$, respectively. }.
\be
< {\bar{B}}_2^{t_R}~ B^{1L}_2 > \sim \Lambda^3_4~~;~~ <
{\bar{B}}^{b_R}_2 B^{2L}_2 > \sim \Lambda^3_4
\ee
These metacolor condensates, combined with the effective residual
interactions (28) of the preon theory, give rise to a form of ETC
interactions.  How these ETC-like interactions arise is sketched
schematically in Fig. 3.  Because the individual $SU(6)$ preon theories
have their own intrinsical dynamical scales, the resulting ETC
interactions need not be the same for top and bottom.  Indeed,
as a result of the metacolor condensate formation one obtains
\begin{eqnarray}
{\cal{L}}_{\rm eff~ETC} &=& \frac{\Lambda^2_4}{\Lambda^2_L
\Lambda^2_{t_{R}}} ~~({\bar{B}}_1^L \gamma_{\mu} \lambda~ B_1^L)
({\bar{B}}_1^{t_R} \gamma^{\mu} \lambda B_1^{t_R}) \nonumber \\
& + & \frac{\Lambda_4^2}{\Lambda^2_L~\Lambda^2_{b_{R}}}~~ ({\bar{B}}^L_1
\gamma_{\mu} \lambda B_1^L )({\bar{B}}_1^{b_R} \gamma^{\mu} \lambda
B_1^{b_R})
\end{eqnarray}
\newpage
\begin{figure}[h]
\vspace{8cm}
\caption[]{Generation of effective ETC interactions, through metacolor
condensation}
\end{figure}

The $B_1$ states, recall, contain both quarks and techniquarks.  Thus
the above
effective ETC interactions, once one turns on the technicolor
interactions, will generate masses for the quarks.  Although the
technicolor condensates are expected to preserve a vectorial $SU(2)$
symmetry~\cite{VW}, the resulting top and bottom quark masses
will be different since the effective ETC interactions for these
states have different strength.  If the scale of the technicolor
condensate is $\Lambda_T$, the one obtains
\be
m_t~\sim~ \Bigl(\frac{\Lambda^3_T \Lambda^2_4}{\Lambda^2_L}\Bigr)
\frac{1}{\Lambda^2_{t_{R}}}~~;~~ m_b~\sim~\Bigl(\frac{\Lambda^3_T
\Lambda^2_4}{\Lambda^2_L}\Bigr) \frac{1}{\Lambda^2_{b_{R}}}
\ee
The masses of the top and bottom quarks will be small compared to the
various preonic dynamical scales $\Lambda_L, \Lambda_{t_{R}}$ and
$\Lambda_{b_{R}}$ - the compositeness scales - provided that these
scales are large compared to the metacolor scale $\Lambda_4$ and the
technicolor scale $\Lambda_T$. If this is so, these states will appear
for all purposes as effectively elementary when probed with energies of
$0(100~GeV)$.
The particular hierarchy between $m_t$
and $m_b$ in the model is due to the difference in the dynamical scales
of the two (right-handed) preon theories, $\Lambda_{t_{R}}$ and
$\Lambda_{b_{R}}$.  These scales typify the physical size of top and bottom
\be
< r_t>~\sim~ \frac{1}{\Lambda_{t_{R}}}~~;~~ < r_b > ~\sim~
\frac{1}{\Lambda_{b_{R}}}~~.
\ee
Thus Eq. (35) contains the interrelation between mass and size alluded
to earlier in Eq. (23).

It is easy to imagine a mechanical extension of this model, so as to
introduce families of quarks \cite{KP}.  To get $N_f$ generations, all
one has to do is to make $N_f$ copies of the $SU(6)^3$ preon model
discussed above.  In this extended model one can reproduce  the
hierarchy of the observed quark mass spectrum by assuming appropriate
dynamical scales for all the various right-handed preon theories..
That is,
\be
m_f~\sim~ \frac{1}{\Lambda^2_{f_{R}}} ~\sim~ < r_f >^2~~.
\ee
However, this naive extension of the toy one family model has many
problems.  These are both of a theoretical and a phenomenological
nature. On the
theoretical side, because this $\Bigl(SU(6)\Bigr)^{3N_f}$
preon theory has so
many preons, at the preon level, all the vectorial interactions - color,
technicolor and metacolor - are {\bf not} asymptotically free.  Thus, it
is no longer clear whether one has control of the dynamics or that the
assumed techniquark and metaquark condensates really occur.  On the
phenomenological side, perhaps even worse disasters occur. The extended
theory \cite{KP}~for $N_f = 3$
 has a natural $[U(1)_V]^3$ family symmetry which, if it
is not broken somehow, prevents generating any quark mixing matrix at
all.  Thus, one has a quark mass hierarchy, but the Cabibbo, Kobayashi
Maskawa matrix $V(_{CKM})$ is identically equal to unity:
\be
V_{CKM} = 1
\ee

It is, perhaps, useful to explain what
prevents quark mixing in the model,
as this is quite generic of
models where generations are obtained by just blindly copying what
happens in one family. For each generation a
vectorial $U(1)$ is preserved in the binding,
under which all preons that carry color have
charge $+1$, and all preons which carry technicolor carry charge
$-1$.  Naively, one would presume that this $[U(1)_V]^3$ symmetry is
also preserved by all condensates. However, because of the chiral nature
of the effective interactions experienced by the $B_1$ and $B_2$ bound
states and because
metacolor itself is probably strong at the compositeness scale,
\footnote{Recall that metacolor is not asymptotically
free in multigenerational models
at the compositeness scale.} it is not clear whether the Vafa - Witten
theorem applies \cite{VW}.  Thus, it is probable that the metacolor
condensates do not respect the $[U(1)_V]^3$ family symmetry.

To get family mixing it is necessary that the metacolor condensates be
not family diagonal
\be < {\bar{B}}^{u_R}_{2i} B^{1L}_{2j}~>~ \sim \Lambda^3_4~
\Sigma^1_{ij}~;~ < {\bar{B}}^{d_R}_{2i}~ B^{2L}_{2j} >~\sim~ \Lambda^3_4
\Sigma^2_{ij}
\ee
with
\be \Sigma^1_{ij} \not= \delta_{ij}~;~ \Sigma^2_{ij} \not= \delta_{ij}
\ee
Unfortunately, this assumption is not enough.
Although these condensates appear to break $[U(1)_V]^3$ to a vectorial
family symmetry $U(1)_V$, the dynamics of the theory is such that there
remains a discrete family symmetry, which ultimately prevents quark
mixing.  To understand this point, consider the analog to  Eq. (34)
which ensues after the formation of the nondiagonal metacolor
condensates given above.  Although these interactions
now connect $B_1$ bound states
of different families, e.g.
\be
{\cal{L}}_{\rm eff~ETC} = \frac{\Lambda^2_4}{\Lambda^2_{Li}~
\Lambda^2_{u_{R}j}}~~ ({\bar{B}}^L_{1i} \gamma^{\mu} \lambda B_{1i}^L)
({\bar{B}}_{ij}^{u_R} \gamma_{\mu}~ \lambda B^{u_R}_{1j} )
\ee
they still preserve the individual $B_{1j}$ - numbers since they always
involve ${\bar{B}}_{1j}~ B_{1j}$ combinations.

To transmit the breaking of family number that occurs through the
metacolor condensates, one needs to make use of the dimension 9
interactions of Eq. (29) which involve only $B_1$
fermions and $B_2$ antifermions
(or vice versa).  Using these interactions, and the metacolor condensates
(39), then indeed one obtains effective interactions which break
explicitly $[U(1)_V]^3$.  Schematically, these interactions take the
form
\be
{\cal{L}}_{B_1~{\rm break}} = \frac{\Lambda^{16}_4}{\Lambda^{14}_{Li}~
\Lambda^5_{u_{R}j},~ \Lambda^5_{d_{R}k}}~ B^L_{1i}~ B^L_{1i} ~ B^L_{1i}
B^L_{1i} {\bar{B}}_{1j}^{u_R} {\bar{B}}^{u_R}_{ij} ~
{\bar{B}}^{d_R}_{1k}~ {\bar{B}}^{d_R}_{1k}
\ee
Unfortunately, although the above interaction breaks the $[U(1)_V]^3$
family symmetry, because of the structure of the theory there remains a
$(Z_2)^3$ discrete family symmetry!  This is a pity because otherwise,
given the high powers of $\Lambda$ involved, the $B_1$ breaking
Lagrangian would really act as a small dynamical perturbation to
generate quark mixing.

Protective family symmetries or their remnants, like the $(Z_2)^3$
symmetry encountered above, are a generic feature of these simple
replica models.  These symmetries prevent quark mixing to occur,
although one
can generate arbitrary diagonal mass matrices.  This problem may
be solved in models where family structure
occurs more naturally.  Given the intriguing interrelation between the
mass of the quarks and their extent, one can well imagine that in a more
realistic model
the largest of the
diagonal elements in the mass matrices of quarks would involve
a $c-t$ transition.  In turn this leads one to speculate that, if there
are FCNC in more realistic theories, these will be most strongly felt
in the $c - t$ channel (and possibly in the $b - s$ channel) \cite{KP}.

Attempts to generate fermion masses in models of dynamical
symmetry breakdown of $SU(2)\times U(1)$ are salutary exercises.  They
serve to remind one that if one
believes that the electroweak theory is broken dynamically, then resolving
the issue of fermion masses is absolutely crucial. Without a a
satisfactory solution to the fermion mass problem, it is difficult to
give credence that $SU(2)\times U(1)$ is broken by some condensate of a
yet to be discovered underlying theory.  Continuing failures to
satisfactorily
resolve
the fermion mass problem give further impetus to the idea
that the electroweak
theory of Glashow, Salam and Weinberg is broken down simply by a scalar
vev.  Before surrendering by default to this idea, however, it seems
worthwhile to continue exploring some of these composite technicolor
models, particularly since the dynamics of combined chiral and vectorial
gauge theories is very rich and could still hide some interesting
surprises.  In doing so, one would be following the example of Salam, who
was not afraid many times in  his career to follow unfashionable
paths, which eventually led to profound later insights.  Whether the
same will occur in this instance remains to be seen.

This work was supported in part by the Department of Energy under grant
No. DE-FG03-91ER40662 TASK C

\end{document}